\documentstyle[12pt,epsf,captions]{article}
\def\singlespace 
{\smallskipamount=3.75pt plus1pt minus1pt
\medskipamount=7.5pt plus2pt minus2pt
\bigskipamount=15pa plus4pa minus4pt \normalbaselineskip=12pt plus0pt
minus0pt \normallineskip=1pt \normallineskiplimit=0pt \jot=3.75pt
{\def\smallskip {\vskip\smallskipamount}} {\def\medskip
{\vskip\medskipamount}} {\def\bigskip {\vskip\bigskipamount}}
{\setbox\strutbox=\hbox{\vrule height10.5pt depth4.5pt width 0pt}}
\parskip 7.5pt \normalbaselines} 

\def\middlespace
{\smallskipamount=5.625pt plus1.5pt minus1.5pt \medskipamount=11.25pt
plus3pt minus3pt \bigskipamount=22.5pt plus6pt minus6pt
\normalbaselineskip=22.5pt plus0pt minus0pt \normallineskip=1pt
\normallineskiplimit=0pt \jot=5.625pt {\def\smallskip
{\vskip\smallskipamount}} {\def\medskip {\vskip\medskipamount}}
{\def\bigskip {\vskip\bigskipamount}} {\setbox\strutbox=\hbox{\vrule
height15.75pt depth6.75pt width 0pt}} \parskip 11.25pt
\normalbaselines} 

\def\doublespace 
{\smallskipamount=7.5pt plus2pt minus2pt \medskipamount=15pt plus4pt
minus4pt \bigskipamount=30pt plus8pt minus8pt \normalbaselineskip=30pt
plus0pt minus0pt
\normallineskip=2pt \normallineskiplimit=0pt \jot=7.5pt
{\def\smallskip {\vskip\smallskipamount}} {\def\medskip
{\vskip\medskipamount}} {\def\bigskip {\vskip\bigskipamount}}
{\setbox\strutbox=\hbox{\vrule height21.0pt depth9.0pt width 0pt}}
\parskip 15.0pt \normalbaselines}

\newcommand{\be}{\begin{equation}}
\newcommand{\ee}{\end{equation}}
\newcommand{\bea}{\begin{eqnarray}}
\newcommand{\eea}{\end{eqnarray}}
      
 \def\a{\alpha}  \def\f{n_f}
\def\d{n_d}    

\begin{document}
\middlespace
\vskip 2cm
\begin{center}
\large {\bf Light adjoint scalars and unification at the string scale.} \\ 
\vskip 1in Mar Bastero-Gil$^{a}$ and Biswajoy Brahmachari$^{b}$ \\
\end{center}
\begin{center}

(a) Scuola Internazionale Superiore di Studi Avanzati \\ 34013
Trieste, ITALY. \\ (b) International Centre For Theoretical Physics,\\
34100 Trieste, ITALY.\\
\end{center}
\vskip 1in
{
\begin{center}
\underbar{Abstract} \\
\end{center}

Following the suggestion of Bachas, Fabre and Yanagida (BFY), we
analyze the gauge coupling unification at the two-loop order, in a
supersymmetric scenario where scalars belonging to the adjoint
representations contribute to the evolution of the couplings from 
intermediate scales onward, and the unification scale is pushed towards the
string scale. Thereafter, we compare the masses of these adjoint scalars 
to the scale at which the hidden sector gauge coupling reaches the
non-perturbative limit at the two-loop order for various possible hidden 
sector gauge groups motivated by the the conjecture of BFY that the 
masses of these adjoint scalars are related to gaugino condensation. We 
also compute the predictions for the top and bottom quark masses in this 
scenario and compare them with those of MSSM. The predicted bottom mass 
improves in the BFY scenario for smaller values of $\alpha_s$.} 

\newpage

The gauge couplings unify at $ M_{string} = 2 \times 10^{16} GeV $ when 
Minimal 
Supersymmetric Standard Model (MSSM) degrees of freedom contributes to 
the evolution of the gauge couplings. However, if the field theory of 
MSSM is a low energy remnant of the string theory, there exists a 
noteworthy discrepancy. This field  theoretic unification scale is 
different from the scale of the string theory, $M_{String}$, which is 
fixed by the intrinsic scale of the string theory, namely $M_{Plank}$ by 
the relation,
\be
M_{string}= g_{string} M_{Plank}.
\ee
One loop string effects could lower this tree level value of the
string scale somewhat, and one calculates \cite{stu,stu1} 
that the string unification scale is modified to,
\be
M_{string}=g_{string} \times 5.27 \times 10^{17}~{\rm GeV} \simeq 
5.27 \times 10^{17}~{\rm GeV},
\ee
where we have assumed $g_{string}=O(1)$. Consequently, the scale
$M_{string}$ is higher than the scale $M_{GUT}$ by a factor of 20
approximately.

In the literature one finds several approaches to rectify the aforementioned
difference between the string scale and the unification scale. There
can be intermediate symmetry groups \cite{int,mar4}, which alter the
running of the gauge couplings and likewise the unification occurs at the
scale of $M_{string}$. Non-standard hypercharge normalizations, which
arise in various string models, can also  accommodate the apparent
mismatch between the string scale and the unification scale
\cite{hyp}.  The heavy string threshold corrections from the string
states \cite{stringth} at the Plank scale or below have been used to reconcile
the mismatch between the string scale and the unification scale.
Non-standard exotic matter \cite{exo} has also been shown to serve
the purpose.

The string models having a $ G \times G $ structure, when broken to
the diagonal subgroup, naturally contains adjoint scalars with zero
hypercharge. Bachas, Fabre and Yanagida (BFY) have pointed out \cite{bfy}
that, if
the mass of these zero-hypercharge adjoint scalars lie in the well
motivated intermediate scale $M_I \sim M^{2/3} m^{1/3}_{susy} \sim
10^{13}$ GeV, the unification scale is pushed up to the string scale
at the one-loop level. In this paper we present a two-loop analysis of
this scenario and find ranges of the masses of the $SU(2)$ and $SU(3)$
adjoint scalars, namely, $M_2$ and $M_3$ for the allowed range of 
$\alpha_s(m_Z)$ \cite{pdg} which gives rise 
to the gauge coupling unification at the scale $M_{string}$. At the
two-loop level there is also a mild variation of the unification scale
to the ratio $\tan \beta = {\langle H_2 \rangle / \langle H_1 \rangle
}$, because at the two-loop level, the running of the gauge couplings
are affected by the simultaneous running of the Yukawa couplings. The
adjoint scalars have no Yukawa couplings with the ordinary Fermions
and so we consider the effect of only the ordinary Yukawa couplings on
the evolution of gauge couplings. 

At the two-loop level the evolution of the gauge couplings are
governed by the equation,
\be
{d \alpha_i \over d t} = {b_i \over 2 \pi}
\alpha^2_i + \sum_{j} {b_{ij} \over 8 \pi^2} \a^2_i \a_j -\sum_{k}
{a_{ik} \over 8 \pi^2} \a^2_i Y_k, \label{2lrg}
\ee
where,
\begin{equation}
b_i= \left(
\begin{array}{c} 2n_f+\frac{3}{5}n_d \\ 
-6+2n_f+n_d+2~\Theta(\mu-M_2) \\ -9+2 n_f + 
3~\Theta(\mu-M_3)\end{array} \right)\,~~~ a_{ik}=
\left(\begin{array}{ccc} \frac{26}{5}&
\frac{14}{5} & \frac{18}{5} \\ 6 & 6 & 2 \\ 4 & 4 & 0
\end{array} \right),
\ee
and,
\be
b_{ij}= \left( \begin{array}{ccc}
\frac{38}{15}n_f+\frac{9}{25}\d & \frac{6}{5}\f+\frac{9}{5}\d &
\frac{88}{15}\f \\ \frac{2}{5}\f+\frac{3}{5}\d & 
-24+14\f+7\d+24~\Theta(\mu - M_2) & 
8\f \\ \frac{11}{5}\f &3\f& -54+\frac{68}{3}\f + 54~\Theta(\mu-M_3)
\end{array} 
\right),
\ee
where, i,j=1,3; k=t,b,$\tau$; $n_d$ are the number of Higgs doublets, $n_f$ 
is the number of Fermion generations, $t=\ln \mu$ and $\Theta(x)=1$, whenever
$x \ge 0$. Fixing the scale $M_X=M_{string}$ we
can find three quantities, from a system of three evolution equations,
using as inputs the quantities $\alpha_1(m_Z)$, $\alpha_2(m_Z)$ and
$\alpha_s(m_Z)$. The evaluated unknowns are $\alpha_G$, $M_2$ and $M_3$. The
calculated values of $M_2$ and $M_3$ are plotted in Figure (1.b). The
magnitude of $M_3$ has an appreciable sensitivity to the input value
of $\alpha_s(m_Z)$. This is also displayed in the Figures (1.a) and (1.b). 

\begin{figure}[htb]
\begin{tabular}{cc}
\epsfysize=8cm \epsfxsize=8cm \hfill \epsfbox{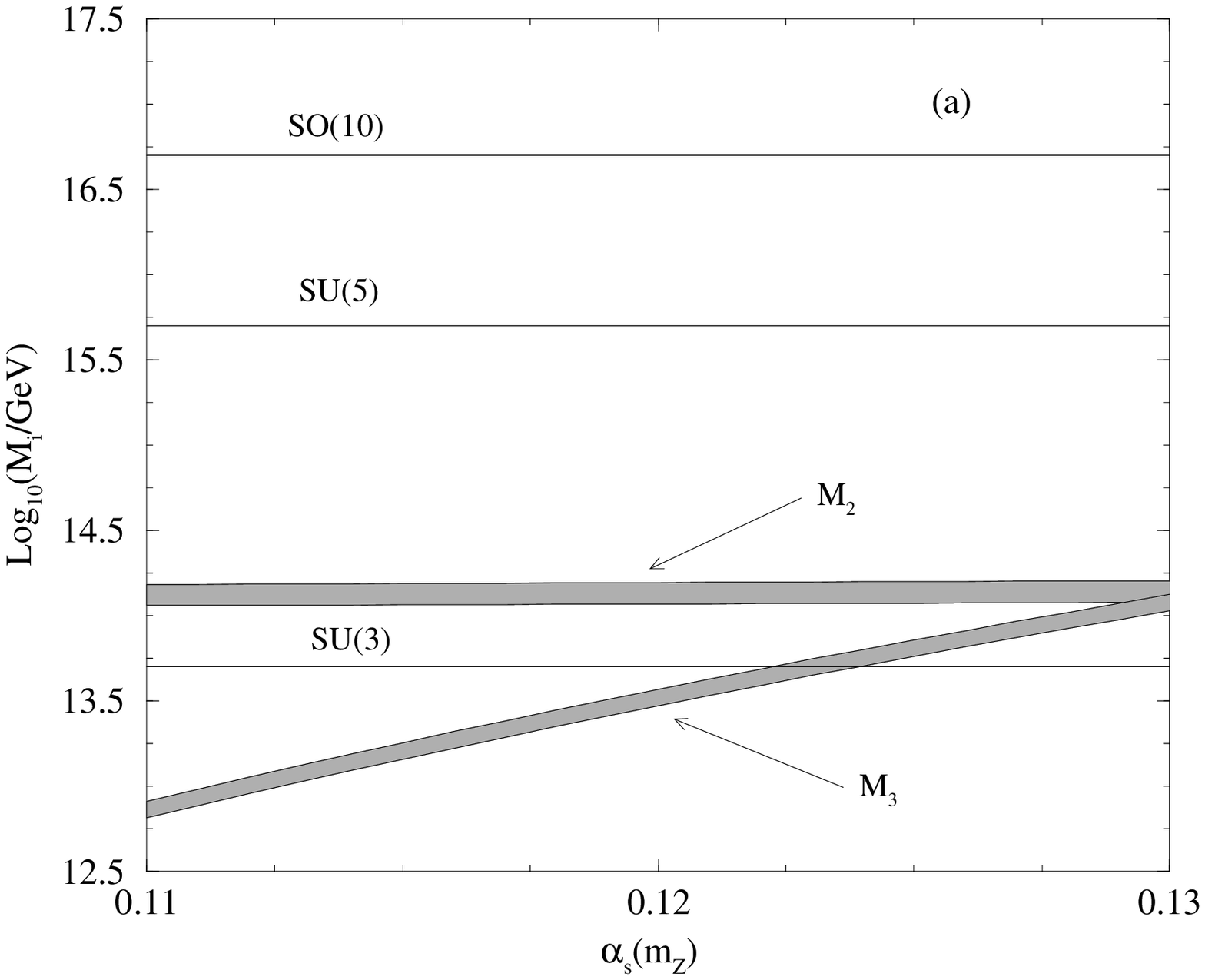} \hfill &
\epsfysize=8cm \epsfxsize=8cm \hfill \epsfbox{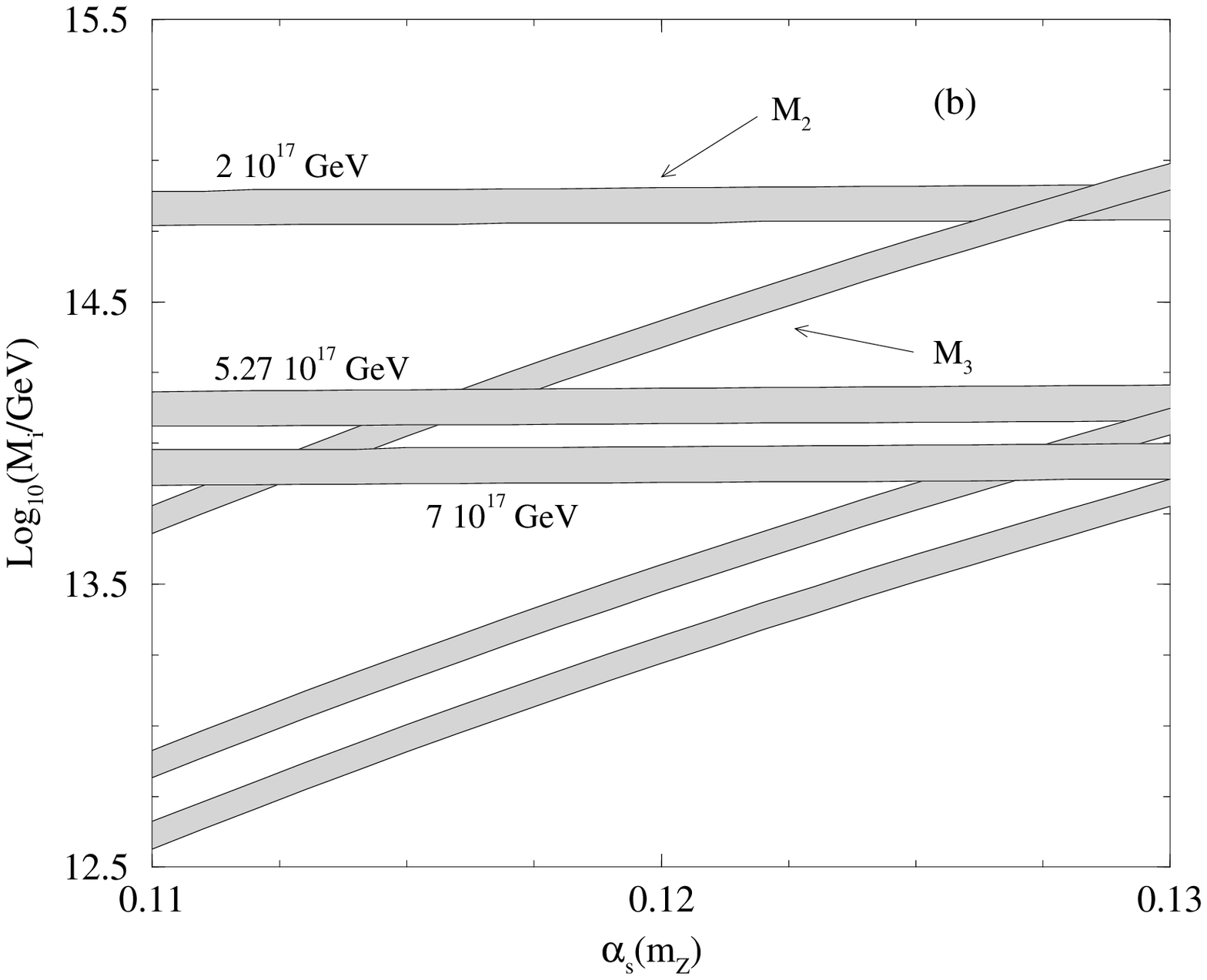} \hfill 
\end{tabular}
\captions{The values of $M_2$ and $M_3$ for various values of the 
unification scale are in Figure (b). The shaded region refers to the 
variation of $\tan \beta$. In Figure (a) we have compared the values of 
$M_2$ and $M_3$ with the condensation scale $M_C$ when $M_{string}$ is 
$5.27 \times 10^{17}$ GeV and $G_H$ is taken as $SU(3)$, $SU(5)$ and 
$SO(10)$.} 
\end{figure}

\begin{figure}[htb]
\begin{tabular}{cc}
\epsfysize=8cm \epsfxsize=8cm \hfill \epsfbox{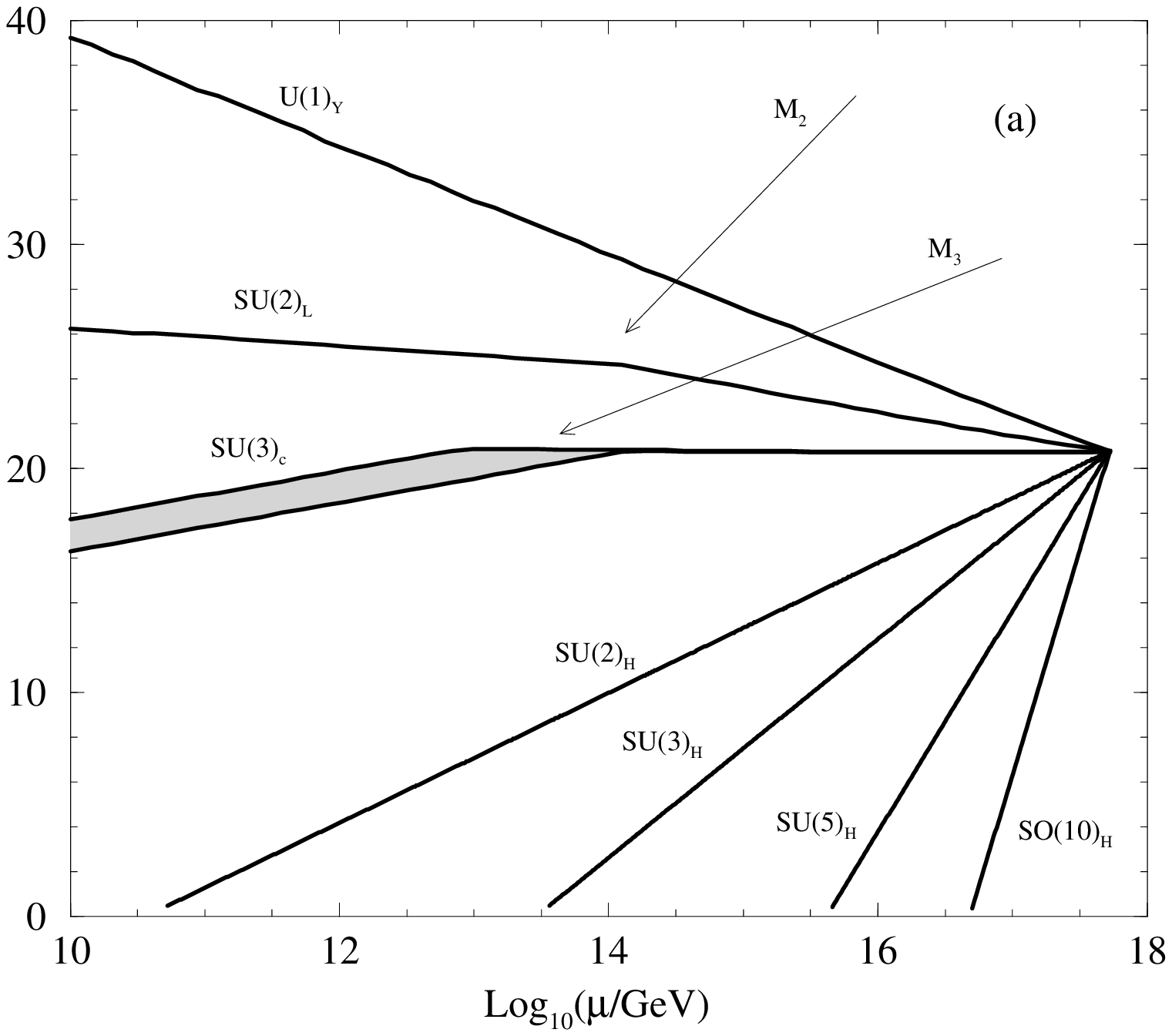} \hfill &
\epsfysize=8cm \epsfxsize=8cm \hfill \epsfbox{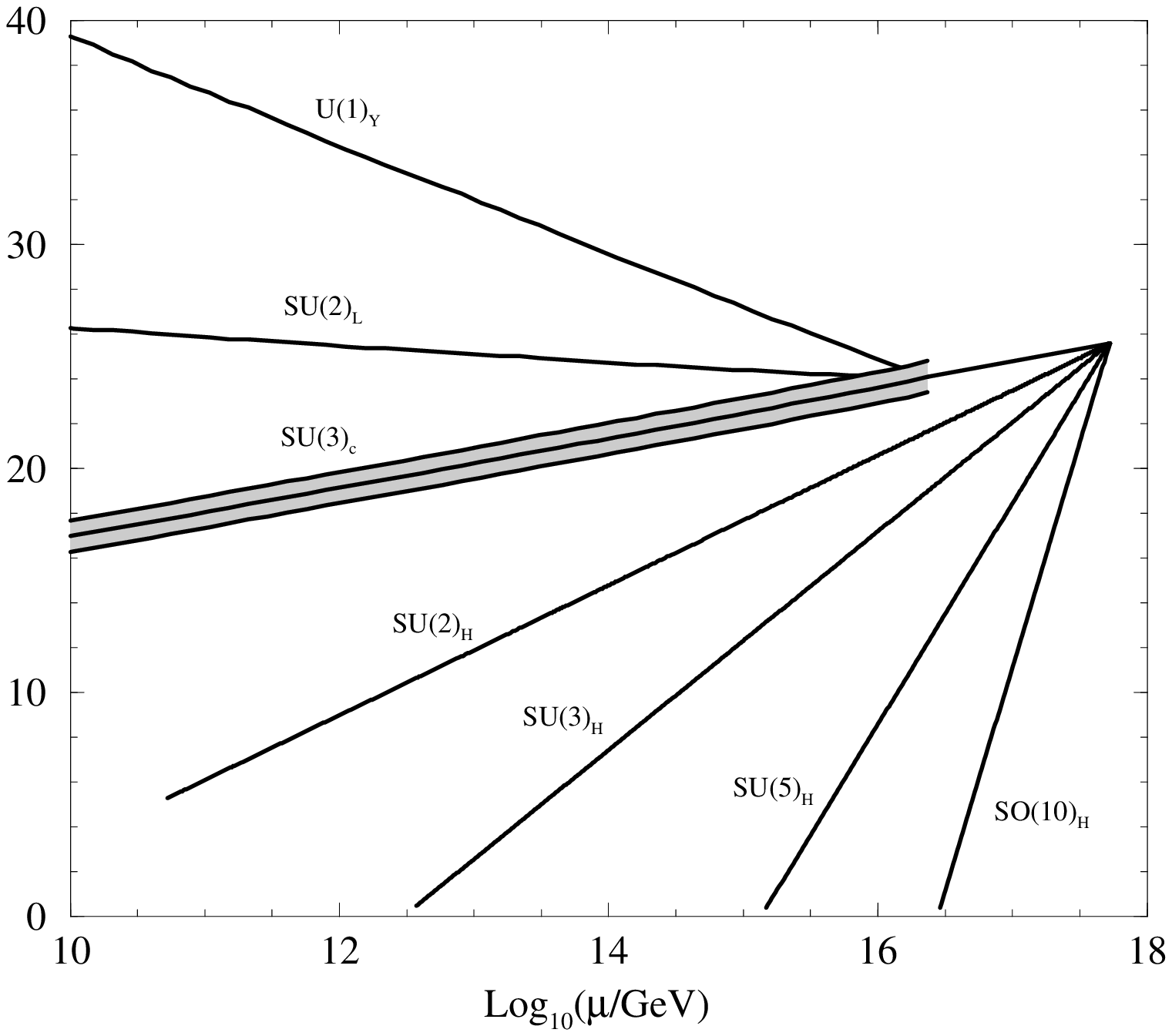} \hfill 
\end{tabular}
\captions{ MSSM unification with a $SU(5)$ GUT with the
minimal Higgs content is Figure (b). The $SU(5)$ coupling at the string 
scale is assumed to be unified with the hidden sector at the string 
scale. Figure (a) is string unification with adjoint moduli at the 
intermediate scale.}
\end{figure}

It is noticeable from Figure (1.b) that the adjoint scalars having the
mass near $10^{12}-10^{14}$ GeV. Numerically, this scale is roughly
near $M^{2/3}_{Plank} m^{1/3}_{susy}$. This is a striking coincidence,
as noted by BFK. Now to check further, we will assume the
unification of the gauge couplings of the hidden and observable
sectors at the scale $M_{string}$. Using the initial value of the
observable sector coupling $\alpha_G(M_{string})$ we evolve the hidden gauge
coupling backwards and search for the scale $M_C$ at which the hidden sector
gauge coupling becomes non-perturbative. We have, for the simplicity,
considered only matter-free gauge groups which are the subgroups of $E_8$, 
eg, $SU(2)$, $SU(3)$, $SU(5)$ or $SO(10)$. In Figure (1.a) we
have compared  the scale $M_C$ with $M_2$ and $M_3$ when $G_H$ is
$SU(3)$, $SU(5)$ and $SO(10)$  respectively. The condensation scale, when 
$G_H$ is $SO(10)$ is large 
compared to the scales $M_2$ and $M_3$. The choice $SU(3)$ as the hidden 
group returns a closer value of $M_C$ to the scales $M_3$ and $M_2$ than 
if $G_H$ is chosen as $SU(5)$. A comment about the results in two-loop 
calculation is in order. Consider the case when $G_H$ is $SU(5)$ and 
$M_{string}= 2 \times 10^{17}$ GeV. In the one loop case the 
scale $M_C$ is $10^{13.5}$ GeV but at the two-loop case this scale becomes 
$10^{15.2}$ GeV. Hence, the two-loop calculation shows a departure from 
the one-loop expectation and brings $M_C$ closer to the intermediate 
scales $M_2$ and $M_3$ when the hidden group is $SU(3)$. This can be seen 
in Figure (2.a), where the running of the couplings are plotted. The 
running of the couplings in the MSSM are plotted in
Figure (2.b) for comparison\footnote{One of the authors B.B  
acknowledges discussions with Q. Shafi on the running of the 
hidden sector gauge coupling.}. We have taken $SU(5)$ as 
the unification group from $M_X \simeq 2 \times 10^{16}$ GeV up to 
$M_{string}$.

\begin{table}[htb]
\begin{center}
\[
\begin{array}{|c|c|c|c|c|}
\hline
G_H&C_2(G) & M_{string}=2 \times 10^{17} & M_{string}= 5.27 \times 10^{17} & 
M_{string}= 7 \times 10^{17} \\ 
\hline SU(2) & 2 & M_C=10^{10.1}\,(10^{7.8}) & M_C=10^{10.9}\,(10^{8.7}) &
M_C= 10^{11.1}\,(10^{9.0}) \\ 
SU(3) & 3 & M_C=10^{13.0} \,(10^{11.0})  & M_C=10^{13.7}\,(10^{11.7})  & 
M_C=10^{13.8}\,(10^{11.9})  \\ 
SU(5) & 5 & M_C=10^{15.2} \,(10^{13.5}) & M_C=10^{15.7} \,(10^{14.1})  & 
M_C=10^{15.9} \,(10^{14.3}) \\ 
SO(10) & 8 & M_C=10^{16.3} \,(10^{14.9}) & M_C=10^{16.7}\,(10^{15.5})  & 
M_C=10^{16.9} \,(10^{15.6}) \\
\hline
\hline
\alpha_s=0.117 & SU(2)~{\rm adjoint}&M_2=10^{14.8}&M_2=10^{14.1}
&M_2=10^{13.9}\\
\alpha_s=0.117 &SU(3)~{\rm adjoint}&M_3=10^{14.2}&M_3=10^{13.4}
&M_3=10^{13.1} \\
\hline
\end{array}
\]
\end{center}
\captions{The masses of the adjoint scalars of $SU(2)$ and $SU(3)$
that leads to gauge coupling unification at the scale $M_{string}$ is in the 
two bottom lines of the table. The values quoted are for $\alpha_s=0.117$. 
It can be compared to the scale $M_C$ at which the hidden sector coupling 
becomes of order one for various gauge groups. For comparison the 
corresponding one-loop values are shown inside the brackets. The gauge 
groups have been designated by their respective quadratic casimirs.} 
\end{table}

\begin{figure}[tbh]
\begin{tabular}{cc}
\epsfysize=8cm \epsfxsize=8cm \hfill \epsfbox{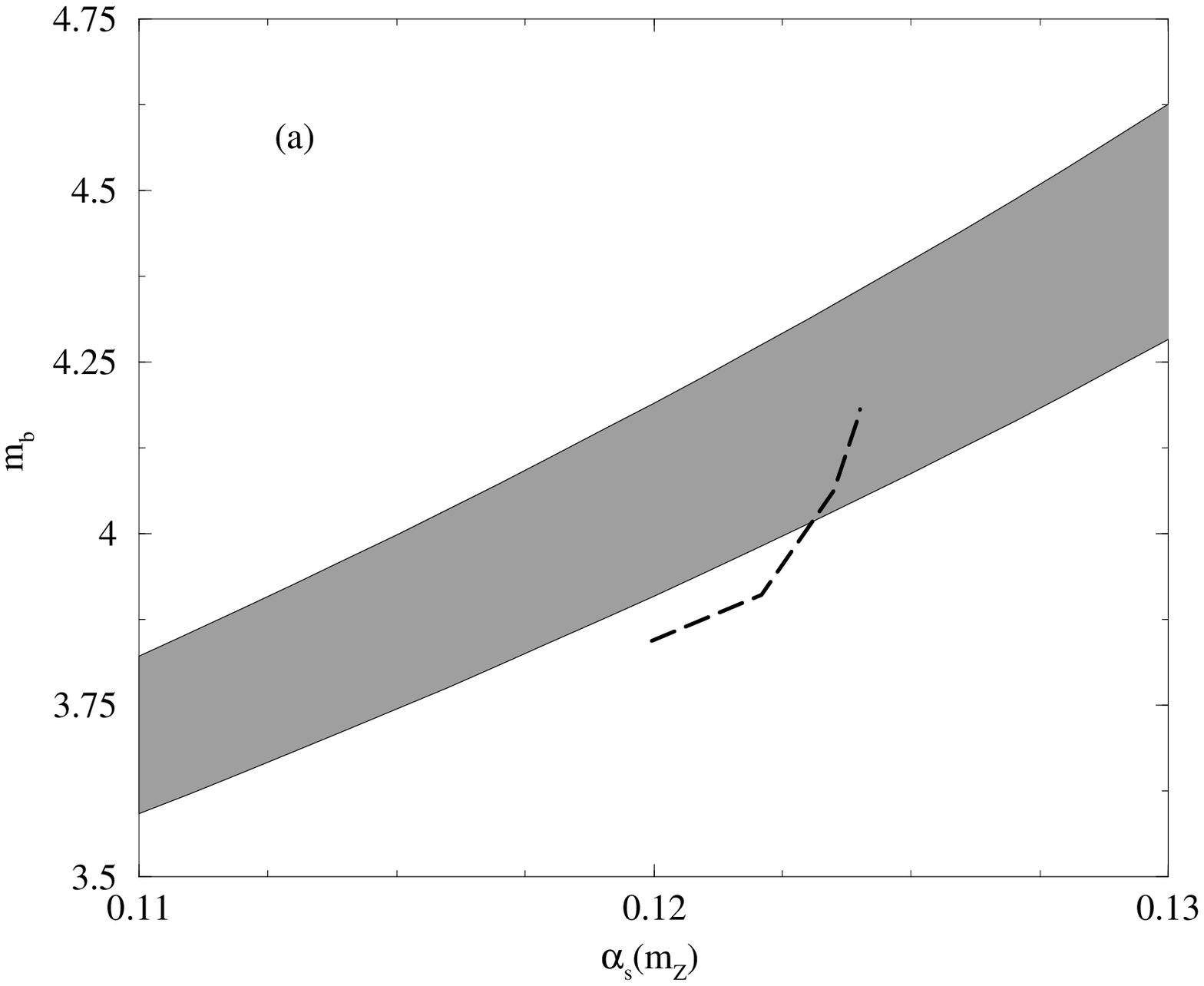} \hfill  &
\epsfysize=8cm \epsfxsize=8cm \hfill \epsfbox{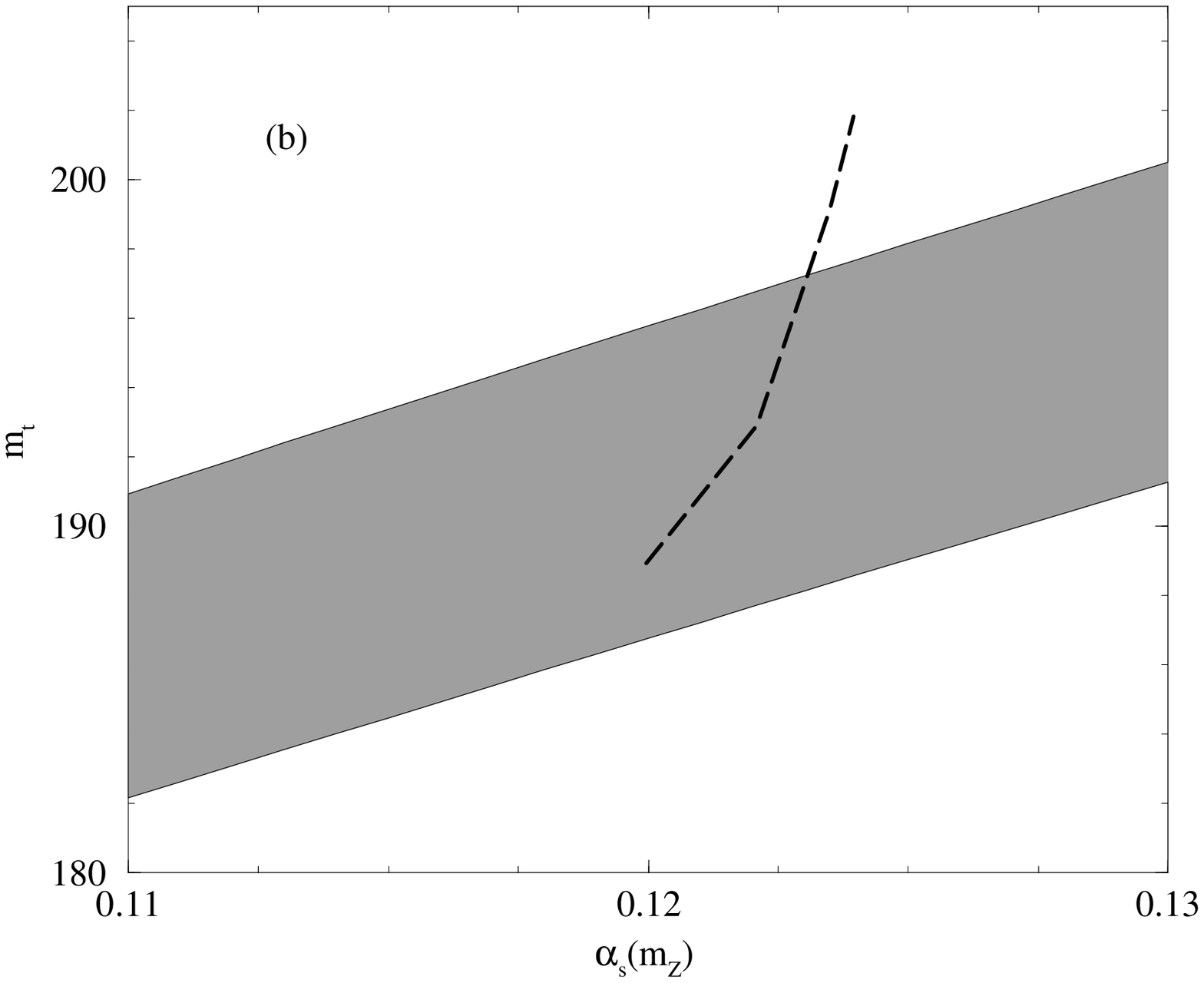} \hfill
\end{tabular}
\captions{The range of the bottom and the top quark mass plotted against 
$\alpha_s$. This range is obtained by varying $\tan \beta$ in the range
approximately 5-60 and $M_{string}$ in the range $2-7 \times 10^{17}$
GeV. Dashed lines are the prediction for the MSSM varying $\tan \beta$
in the same range.}
\end{figure}

Before we conclude, we give some results of a two-loop Yukawa coupling
analysis of this model. The relevant beta function coefficients are 
summarized in Ref \cite{mar4}. The Top quark mass is 
determined from the quasi-infrared fixed point of the top quark Yukawa 
coupling \cite{2l}. The presence of the adjoints alter the running of the 
gauge couplings above the intermediate scale which has secondary effects 
on the running of the Yukawa couplings. Even though we do not expect a 
large change in the 
individual Yukawa couplings at low energy, the ratios of the Yukawa 
couplings are sensitive to the presence of these extra matter. To check 
these issues, we have plotted the prediction of the top quark and the 
bottom quark masses in the Figures (3.a) and (3.b). The bottom quark mass 
can be calculated under the well-known assumption of b-$\tau$ 
unification\footnote{b-$\tau$ unification is natural in GUT models
like $SU(5)$ or $SO(10)$. Here, we consider a similar scenario in
string unification.}. The prediction of the bottom quark mass, having been 
calculated from the $ratio$ of the bottom and $\tau$ Yukawa coupling, 
differs notably when the extra adjoint matter is present. To see this we 
have compared the present case with that of MSSM in Figure (3.a) and (3.b) 
in dashed 
lines. The bands in the figures (3.a) and (3.b) are obtained by 
varying $\tan \beta$ in the range 5-60 and $M_{string}$ in the 
range $2-7 \times 10^{17}$ GeV. In MSSM the unification occurs for 
only a narrow 
region in the $\alpha_s$ space, whereas, the presence of adjoint moduli 
opens up the
parameter space further and as an welcome result unification occurs
for a larger range in $\alpha_s$. This fact is reflected in Figure (3)
where the prediction of the top and bottom quark masses are displayed
as narrow lines, whereas in the present case with adjoint moduli the
predictions are wide bands. The bottom quark mass can be within the
experimental range for lower values of $\alpha_s$, for which the
prediction in the MSSM case is unavailable due to the absence of gauge
coupling unification and hence b-$\tau$ unification.
In Table 2 we have compared the predictions of $m_t^{pole}$, $m_b^{pole}$ 
and $\tan \beta$ in the MSSM with those obtained in the presence of 
adjoints for the required value of $\alpha_s(m_Z)$ which gives rise to 
unification in MSSM [Note that this specific value of $\alpha_s$ 
depends on $Y_b$ or equivalently on $\tan \beta$]. Typically, the value 
of $m_t^{pole}$ diminishes while $m_b^{pole}$ increases slightly in the 
BFY scenario.  

\begin{table}[htb]
\begin{center}
\[
\begin{array}{|c||c||c||c||c|}
\hline
Y_t(M_{string}),Y_b(M_{string})& \alpha_s(m_Z) & m^{pole}_t & m^{pole}_b & 
\tan \beta\\ && 
\begin{array}{cc} 
\hline 
MSSM & Moduli 
\end{array}
& \begin{array}{cc}
\hline
MSSM & Moduli 
\end{array} 
&
\begin{array}{cc} 
\hline 
MSSM & Moduli 
\end{array}
\\
\hline
1,1^{~~~~}& 0.120&     \begin{array}{c|c} 189  &  187  \end{array}& 
	      \begin{array}{c|c} 3.84 &  3.94 \end{array}& 
	      \begin{array}{c|c} 63.2 & 61.0  \end{array} \\
1,10^{-1}& 0.122& \begin{array}{c|c} 193  & 190   \end{array}& 
	      \begin{array}{c|c} 3.91 &  4.00 \end{array}& 
	      \begin{array}{c|c} 59.1 & 57.7  \end{array}\\
1,10^{-2}&0.123& \begin{array}{c|c}  199  &  197  \end{array}& 
	     \begin{array}{c|c} 4.06  &  4.18 \end{array}& 
	     \begin{array}{c|c}  39.9 &  40.6 \end{array}\\
1,10^{-3}&0.124& \begin{array}{c|c}  202  & 200   \end{array}& 
	     \begin{array}{c|c} 4.16  &  4.31 \end{array}& 
	     \begin{array}{c|c}  15.8 &  16.5 \end{array}\\
1,10^{-4}&0.124& \begin{array}{c|c}  199  & 197   \end{array}& 
	     \begin{array}{c|c} 4.18  &  4.33 \end{array}& 
	     \begin{array}{c|c}   5.1 &   5.3 \end{array}\\
\hline
\end{array}
\]
\end{center}
\captions{The predictions of $m^{pole}_t$ and $m^{pole}_b$ in MSSM 
have been compared with those of BFY scenario for $\alpha_s(m_Z)$ which 
gives rise to unification in the MSSM case. For the BFY scenario 
$M_{string}=5.27 \times 10^{17}$. $Y_b=h^2_b/4\pi$, where $h_b$ is the 
bottom quark Yukawa coupling. The low energy value of $\tan \beta$ has 
been calculated using $Y_\tau(m_\tau)$ from the RGE 
and $m_\tau=1.777$ GeV; and this value of $\tan \beta$ has been used to
estimate the top and the bottom masses.}
\end{table}
 
In conclusion, we have computed the two-loop running of the gauge
couplings in a supersymmetric scenario with extra adjoint matter from
intermediate scales onwards. Setting the unification scale at the
string scale, the masses of these adjoints and the unification
coupling have been calculated. The masses of the adjoints turn out to be 
of the same order as that of the expected gaugino condensation scale 
$M^{2/3}_P m^{1/3}_{susy}$ in some cases as displayed in Figure (1) and 
Table 1. Taking the gauge coupling $\alpha_H$ for the hidden sector 
group $G_H$, 
to be equal to the unification coupling at the scale $M_{string}$, we 
have evolved back this value to the scale $M_C$, where $\alpha_H$  
becomes non-perturbative at the two-loop order, for different choices  
of the hidden (matter free) group $G_H$. The two-loop running of the hidden 
sector is called for, because the scale $M_C$ differs from the one-loop 
expectation (shown inside brakets in Table-1) by more than one order 
of magnitude. When $G_H$ is $SU(3)$ or $SU(5)$ the scale $M_C$ works out 
to be of the order of the masses of $M_2$ and $M_3$. In comparison to 
MSSM, unification occurs for wider range of $\alpha_s$. The bottom quark 
pole mass prediction from 
b-$\tau$ unification improves compared to that of MSSM for smaller values 
of $\alpha_s$ like 0.115. The top quark mass is evaluated assuming that 
the top quark Yukawa coupling at $M_{string}$ is within domain of the 
quasi-infrared fixed point at the scale $m_{top}$. The range for the 
prediction of the top quark mass is consistent with experimental numbers 
in the moduli scenario.   

We thank C. Bachas and Q. Shafi for critical comments and discussions.

\newpage
\doublespace 

\end{document}